\documentclass[a4paper,twoside,10pt]{article}
\usepackage{graphicx,publaob}

\def\papertype{Progress Report}
\begin{document}

\title{SPECTRAL CHARACTERISTICS OF FAST ROTATING METAL-POOR MASSIVE STARS}

\authors{B. Kub\'atov\'a$^{1}$, D. Sz\'ecsi $^{2}$}

\address{$^1$Astronomical Institute of the Czech Academy of Science, Fri\v cova 298, 251 65 Ond\v rejov, Czech Republic}
\Email{brankica.kubatova}{asu.cas}{cz}
\address{$^2$Institute of Astronomy, Faculty of Physics, Astronomy and Informatics, Nicolaus Copernicus University, Grudziadzka 5, 87-100 Toru\'n, Poland}

\markboth{\runningfont SPECTRAL CHARACTERISTICS OF FAST ROTATING METAL-POOR MASSIVE STARS}
{\runningfont B. KUB\'ATOV\'A,
\runningit{\& D. SZ\'ECSI}}

\vskip-1mm

\abstract{Low-metallicity massive stars are assumed to be progenitors of certain supernovae, gamma-ray bursts, and gravitational wave emitting mergers. These exotic phenomena contribute to their host galaxies through strong ionizing radiation and mechanical feedback. Here we investigate a certain type of very metal-poor (0.02 $Z_{\odot}$) hot massive single stars that rotate fast and evolve chemically homogeneously. Combining state-of-the-art theories of stellar evolution and stellar atmospheres modelling we predict synthetic spectra of these stars corresponding to different masses and evolutionary phases. The predicted spectra in early evolutionary phases is classified mainly as an early-O type giant or supergiant while in later evolutionary phases most of our model spectra are assigned to the WO-type spectral class. The Hubble Space Telescope’s (HST) Ultraviolet Legacy Library of Young Stars as Essential Standards (ULLYSES) program will enable us to compare our predicted  spectra with observations of stars of similar nature (e.g., metal-poor stars in \textit{Sextant~A}).
}

\section{INTRODUCTION}

Low-metallicity massive stars are essential for understanding Cosmic history. It is assumed that early metal-poor/metal-free stars were sufficiently hot to re-ionize the early Universe (see e.g., Abel et al., 2002). These stars also might have been contributing to our Galaxy’s and other galaxies' chemical compositions. Therefore, spectroscopic observations of these stars and their analyses would be of great importance to better understand their nature and to quantify the feedback of these stars to their host galaxies. However, due to our insufficient observing capacities we are still unable to observe the first and second generations of stars in the early Universe. Even extremely metal-poor stars like those in the local star-forming dwarf galaxies (e.g., \textit{Sextant~A}) are rarely observed and analysed as individual objects (see e.g., Garcia et al, 2019). 
The spectral appearance of stars with metallicity below $0.1\,Z_{\odot}$ is still unknown. 

To better understand the nature of very metal-poor stars, Sz\'ecsi et al. (2015) calculated stellar
evolutionary models of single massive stars with the composition of \textit{I~Zw~18}, that is, a metallicity of $Z\sim 0.02\,Z_{\odot}$. They found that the models that rotate fast evolve chemically homogeneously due to rotational mixing (Maeder, 1987). They named them Transparent Wind UV-Intense (TWUIN) stars (see, e.g., Sz\'ecsi, 2016) because of the predicted weak, optically thin stellar winds combined with being hot, and thus emitting most of their radiation in the ultraviolet (UV) band. In the Hertzsprung–Russell (HR) diagram these stars evolve blue-wards during the core-hydrogen-burning (CHB) phase, unlike the non-rotating and slow-rotating stars which evolve to the right side of the HR diagram as metal-rich massive stars in our Galaxy do (see Fig. 5 in Sz\'ecsi et al. 2015).

So far TWUIN stars have only been predicted by current state-of-the-art stellar evolutionary physics. To examine the spectroscopic nature of these stars and to quantify their contribution to \textit{I~Zw~18} or other similarly metal-poor galaxies, we study their spectral appearance, explore expected observable features, and classify these stars accordingly.

\section{STELLAR EVOLUTIONARY MODELS}

Using the ‘Bonn’ stellar evolution code we calculated three evolutionary tracks of fast-rotating massive single stars of very low-metallicity, $Z\sim0.02\,Z_{\odot}$ (see Sz\'ecsi et al., 2015 and Sz\'ecsi, 2016), with initial masses (M$_\mathrm{ini}$) of 20, 59, and 131~M$_{\odot}$, and with initial rotational velocities of 450 km/s, 300 km/s, and 600 km/s, respectively. For each M$_\mathrm{ini}$ we calculated five models. Four models correspond to CHB phases with surface helium mass fractions (Y$_\mathrm{S}$) of 0.28, 0.5, 0.75, and 0.98. The fifth model corresponds to the central-helium-burning (CHeB) phase with a central helium mass fraction (Y$_\mathrm{C}$) of 0.1. For more details about main model parameters and evolutionary tracks see Table 1 and Fig. 1 in  Kub\'atov\'a et al. (2019).

Different mass-loss rate prescriptions were used for different evolutionary stages. The Vink mass-loss rate prescription for hot O-type stars (see Eq.~24 in Vink et al., 2001) was applied when Y$_\mathrm{S}$ was
lower than 0.55 while for phases when Y$_\mathrm{S}$ was higher then 0.7, the WR-type prescription of Hamann et al. (1995) reduced by a factor of 10 (as suggested by Yoon et al., 2006) was used. During the whole CHeB phase, the WR-type prescription was applied everywhere. To account for uncertainties in the mass-loss rate predictions for every model we calculated two sets of synthetic spectra for two different values of mass-loss rates, one with a nominal mass-loss rate as implemented in the evolutionary models, and the other with a reduced value mass-loss rate that was a factor 100 lower than the nominal value. Choosing a factor of 100 is motivated by the fact that metallicity-dependence of WR winds can be steeper (see Hainich et al., 2015). Applied values of nominal and reduced mass-loss rates used in the synthetic spectra computations can be found in Table 2 in  Kub\'atov\'a et al. (2019).

\begin{figure}[t]
\centering
\includegraphics[width=1.\textwidth]{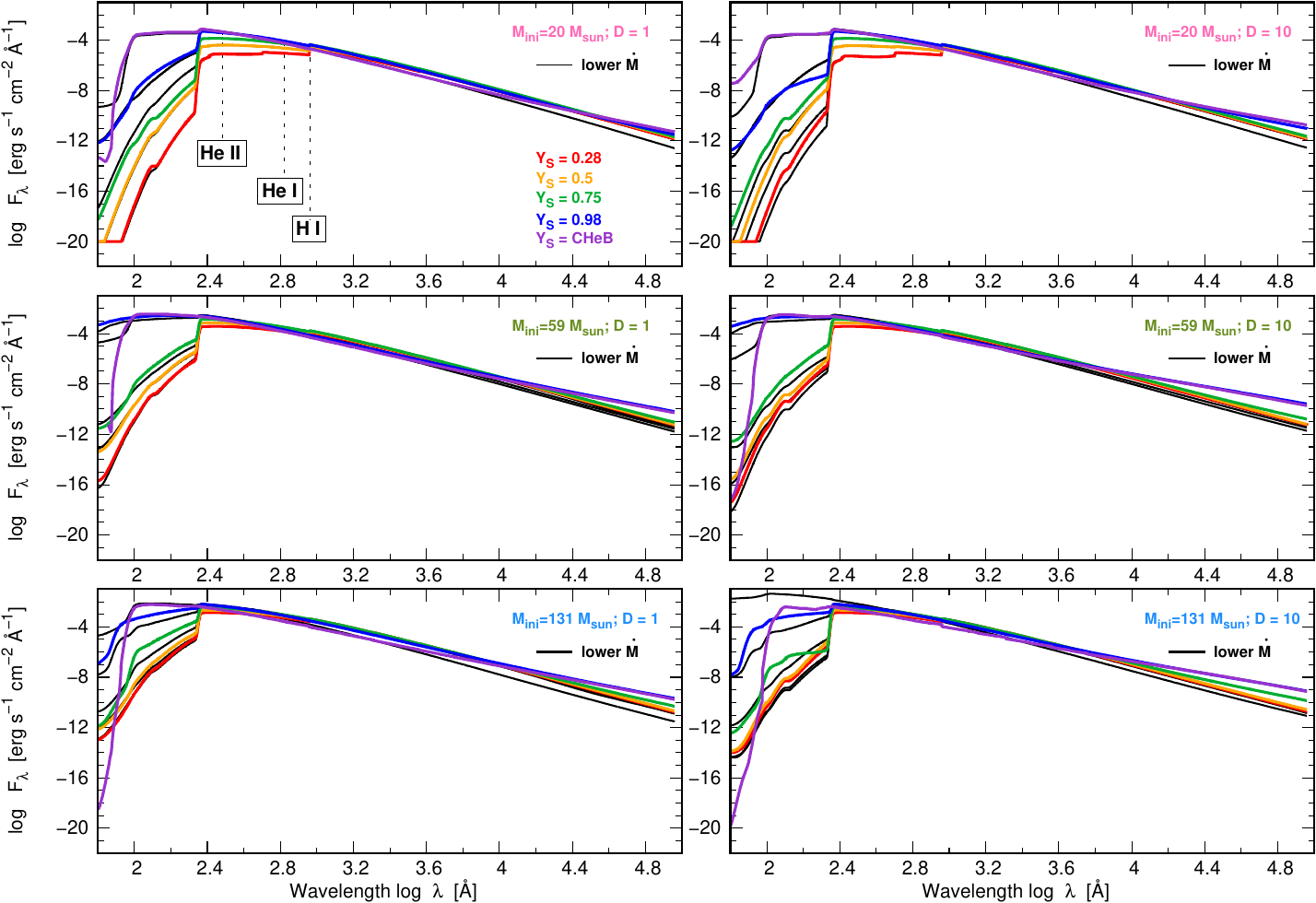}
\caption{The continuum SED of chemically-homogeneously evolving star with different M$_\mathrm{ini}$ (marked in the right-top corner of each panel), and in different evolutionary stages marked by the value Y$_\mathrm{S}$ in the top-left panel (colored lines). Models were calculated assuming nominal values of the mass-loss rate (i.e., the values used in the stellar evolutionary models) and a smooth wind (D=1). Black lines represent the SEDs of the same models but with 100 times lower mass-loss rate. H~{\sc i}, He~{\sc i}, are He~{\sc ii} represent ionisation stages of hydrogen and helium, respectively.}
\label{fig:SED}
\end{figure}

\section{STELLAR ATMOSPHERE AND WIND MODELS}

To calculate synthetic spectra we used the Potsdam Wolf-Rayet ({\tt{PoWR}}) stellar atmosphere code. Stellar parameters required as an input in the {\tt{PoWR}} code (the stellar temperature $T_{*}$, the stellar mass $M_{*}$, and the stellar luminosity $L_{*}$) were adopted from the stellar evolutionary model sequences (see Table 1 in Kub\'atov\'a et al., 2019). Details about the assumptions and numerical methods of model atmosphere calculations of the  {\tt{PoWR}} code can be found in  Hamann \& Gr\"afener (2004) and Sander et al. (2015). 

To ensure that all sources that significantly contribute to opacity were taken into account, all relevant elements (H, He, C, N, O, Ne, Mg, Al, Si, and Fe) are used in the spectra calculation. Abundances of these elements were adopted from the stellar evolutionary model sequences. Additional elements which were not included in the stellar evolutionary models such as P, S, Cl, Ar, K, and Cl (with abundances of $Z_{\odot}$/50) were also included in the {\tt{PoWR}} code to account for their potential contribution to the wind driving. 

With a given value of mass-loss rate as it was used in the stellar evolutionary models, the density stratification $\rho$(r) in the wind was calculated via the continuity equation, $\dot{M}=4\pi r^{2}v(r)\rho(r)$. The velocity field was prescribed by the so-called $\beta$-law, $v(r)=v_{\infty}(1-R_{*}/r)^\beta$ (see, e.g., Lamers \& Cassinelli, 1999). For $\beta$ parameter we assumed values of 0.8 or 1.0 and for $v_{\infty}$ we assumed 1000~km/s for all models. The wind inhomogeneities (i.e., clumping) were treated in the microclumping approximation (see Hamann \& Koesterke, 1998). All clumps were assumed to be optically thin, the inter-clump medium was assumed to be void, and the clumping factor (D) was assumed to depend on radius (for more details see e.g., Sander et al. 2017). In order to examine the importance of clumping on the spectral appearance of metal-poor massive stars, we calculated two sets of synthetic spectra for two different values of the clumping factor, one for D=1 (smooth wind), and another one for D=10. 

\section{RESULTS}

\begin{figure}[t]
\centering
\includegraphics[width=0.49\textwidth]{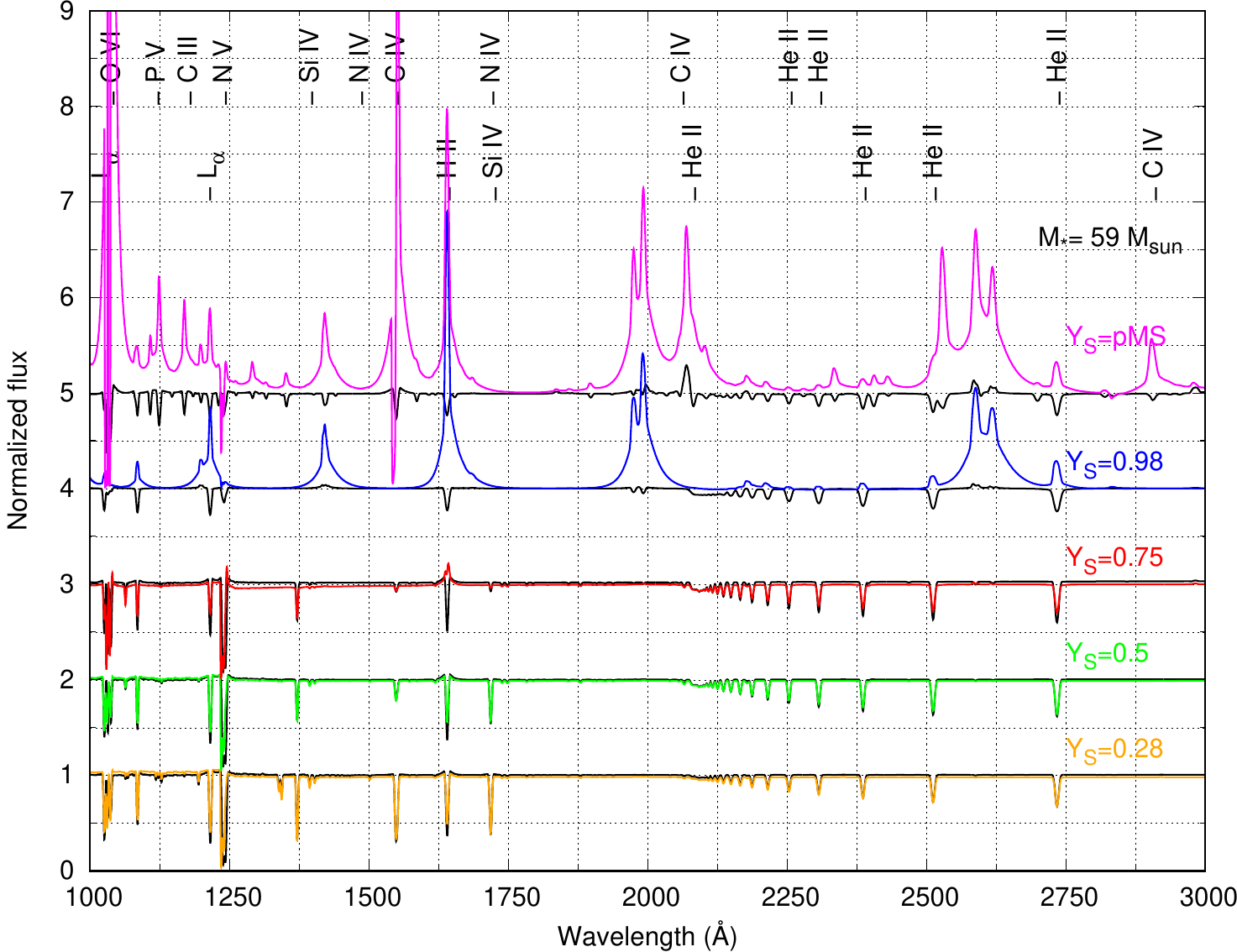}
\includegraphics[width=0.49\textwidth]{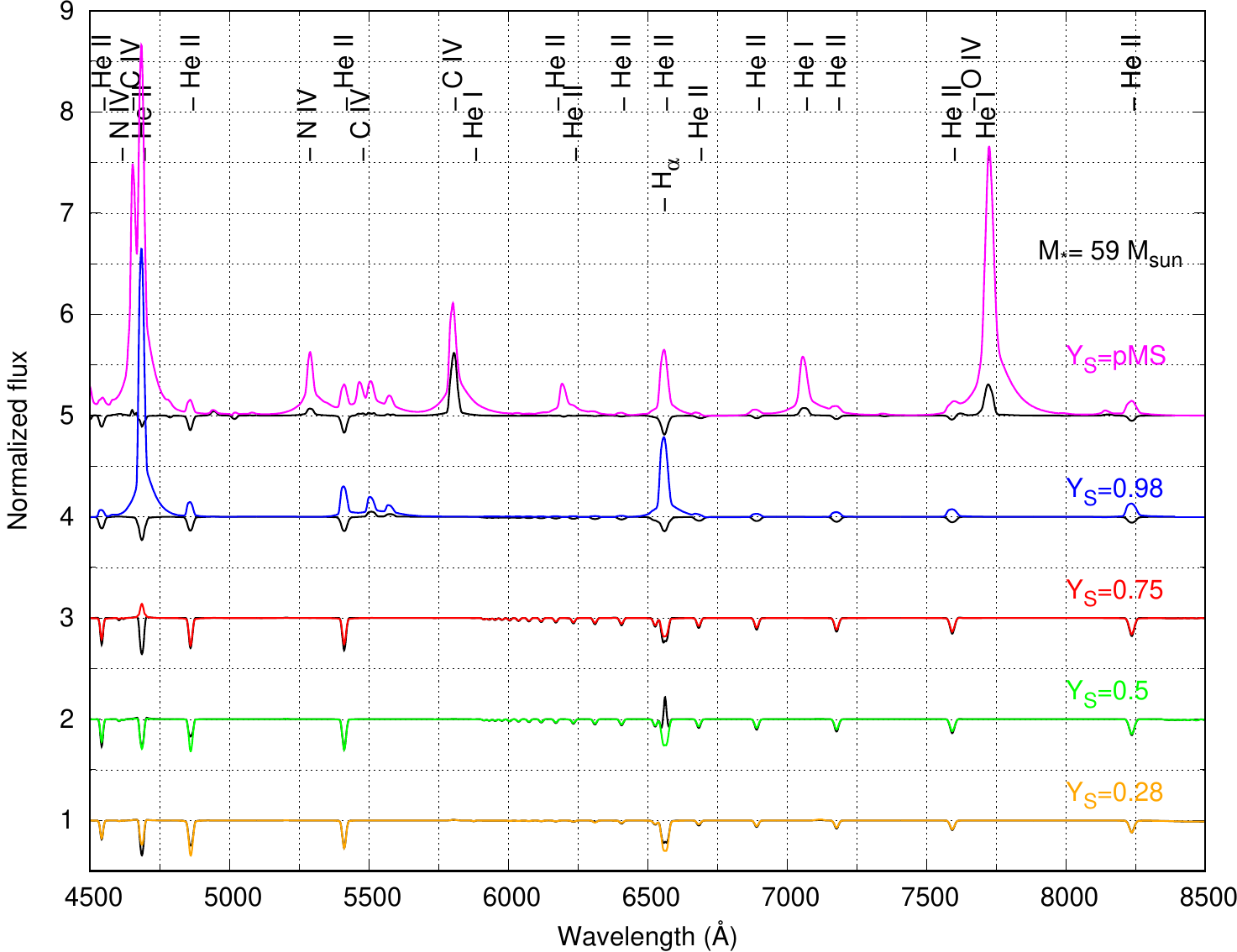}
\caption{Synthetic spectra of chemically-homogeneously evolving stars in the UV (left panel) and  optical (right  panel) regions with M$_\mathrm{ini}=59~M_{\odot}$ and in different evolutionary phases marked by the value $Y_{\mathrm{S}}$. Colored lines represent models with nominal mass-loss rates (i.e., the values used in the stellar evolutionary models) and smooth wind while the corresponding black lines represent the same models but with 100 times lower mass-loss rates.}
\label{fig:OPT-UV}\vspace{-3mm}
\end{figure}

In total we calculated 60 models of chemically-homogeneously evolving stars for three different values of M$_\mathrm{ini}$, for five different evolutionary stages defined by $Y_{\mathrm{S}}$, and for two different values of mass-loss rates and wind clumping. The line profiles were calculated taking line broadening due to radiation damping, pressure broadening, and rotational broadening into account.

\subsection{Spectral characteristics}

For each model we calculated a spectral energy distribution (SED) and found that the maximum of radiation emitted by chemically-homogeneously evolving stars is in the far- and extreme UV region and that the total emitted flux is not very sensitive to variations of either the mass-loss rate or clumping. Smaller differences between SEDs can be seen at wavelengths
shorter than 227~{\AA} and longer than 10~000~{\AA} and in the evolutionary stages shortly before the end of the CHB ($Y_{S}=0.98$) phase and in the post-main sequence (CHeB) phase (see Fig.~\ref{fig:SED}). The SEDs reveal that the radiation with frequencies higher
than the H\,{\sc i}, He\,{\sc i}, and He\,{\sc ii} ionization limits increases both with the initial mass and during the evolution of the stars. More massive and more evolved stars emit more ionizing flux.

The synthetic spectra of models in earlier evolutionary phases ($Y_{S}~<~0.5$) show almost exclusively absorption lines regardless of their M$_\mathrm{ini}$. In later evolutionary phases lines start to appear in emission while in the CHeB phase almost all lines are found in emission that corresponds to a Wolf--Rayet (WR) type stellar spectrum.
The helium emission lines are found to be strong (see Fig.~\ref{fig:OPT-UV}).

From the spectral characteristics it can be inferred that chemically-homogeneously evolving stars in early evolutionary phases can be assigned to the term TWUIN~star since they show spectral features that are typical of weak and optically thin winds. However, in later evolutionary phases, these stars begin to exhibit spectral features that are typical for WR~stars with strong and optically thick winds (see purple lines in Fig.~\ref{fig:OPT-UV}).

\subsection{Spectral classification}

Using the Morgan–Keenan spectroscopic classification scheme, we assigned spectral classes to all the synthetic spectra of chemically-homogeneously evolving stars presented above. 

We found that the models of these stars during the CHB phase emit spectra which can be assigned to the spectral class O4 or earlier with luminosity classes I (a supergiant), III (a giant), or V (a dwarf). The spectra show strong He~{\sc ii} emission lines without almost any metal line. During the CHB phase these stars are predicted to be much hotter than the O-type stars that have been observed spectroscopically so far (down to 0.1 $Z_\odot$). 

Later during the CHeB phase, we found that chemically-homogeneous evolution leads to WO type stars (i.e., WR stars in which ionised oxygen dominates in the spectrum) with spectra completely without nitrogen lines. For more details of the spectral classification see Table 4 and Appendix A in Kub\'atov\'a et al, (2019).

\section{SUMMARY}

Combining stellar evolution and atmospheric modeling we predicted the spectral appearance of chemically-homogeneously evolving stars. We showed that in the early evolutionary phases chemically-homogeneously evolving stars exhibit spectral features that are typical of weak and optically thin winds. Thus the term TWUIN indeed applies to them. In evolved phases we predicted that these stars begin to develop spectral features that are typical of WR stars with strong and optically thick winds. 

Our spectral predictions reveal that an extremely hot, early-O type star observed in a low-metallicity galaxy could be the result of chemically homogeneous
evolution and might therefore be the progenitor of a long-duration gamma-ray burst or a type Ic supernova. If born in a binary, the system may even evolve to become a double-compact object merger leading to the emission of gravitational waves (Marchant et al, 2016).
TWUIN stars may play an important role in re-ionizing the Universe because they are hot without showing prominent emission lines during most of their lifetimes.

\vskip2mm

\centerline{\bf Acknowledgements}

\vskip2mm

\noindent This work was partly supported by a grant GA ČR 22-34467S. The Astronomical Institute Ond\v rejov is supported by the project RVO: 67985815. D.Sz. was funded in part by the National Science
Center (NCN), Poland under grant number OPUS 2021/41/B/ST9/00757. For the purpose of Open Access, the author has applied a CC-BY public copyright license to any Author Accepted Manuscript (AAM) version arising from this submission.

\vskip-.5cm

\references

Abel, T., Bryan, G. L., Norman, M. L., 2002, \journal{Science}, \vol{295}, 5552.

Garcia, M., Herrero, A., Najarro, F., Camacho, I., Lorenzo, M.: 2019, \journal{Monthly Notices of the Royal Astronomical Society}, \vol{484}, 1.

Hainich, R., Pasemann, D., Todt, H., Shenar, T., Sander, A., Hamann, W.-R.: 2015, \journal{Astronomy \& Astrophysics}, \vol{581}, A21.

Hamann, W.-R., \& Koesterke, L.: 1998, \journal{Astronomy \& Astrophysics}, \vol{335}, 1003.

Hamann, W.-R., Koesterke, L., Wessolowski, U.: 1995, \journal{Astronomy \& Astrophysics}, \vol{299}, 151.

Hamann, W.-R. \& Gräfener, G. 2004, \journal{Astronomy \& Astrophysics}, \vol{427}, 697.

Kub\'atov\'a, B., Sz\'ecsi, D., Sander, A. A. C., Kub\'at, J., Tramper, F., Krti\v cka, J., Kehrig C., Hamann, W.-R., Hainich, R., and T. Shenar: 2019, \journal{Astronomy \& Astrophysics}, \vol{623}, A8.

Lamers, H., \& Cassinelli, J.: 1999, \journal{Introduction to Stellar Winds} (Cambridge University Press).

Maeder, A.: 1987, \journal{Astronomy \& Astrophysics}, 178.

Sander, A., Shenar, T., Hainich1, R., Gímenez-García, A., Todt, H., Hamann, W.-R.: 2015, \journal{Astronomy \& Astrophysics}, \vol{577}, A13.

Sander, A. A. C., Hamann, W.-R., Todt, H., Hainich, R., Shenar, T.: 2017, \journal{Astronomy \& Astrophysics}, \vol{60}, A86.

Sz\'ecsi, D., Langer, N., Sanyal, D., de Mink, S., Evans, C.~J., Dermine, T.: 2015, \journal{Astronomy \& Astrophysics}, \vol{581}, A15.

Sz\'ecsi, D.: 2016, \journal{PhD thesis, Mathematisch-Naturwissenschaftlichen Fakultät der Universität Bonn}.

Vink, J., de Koter, A., Lamers, H.: 2001, \journal{Astronomy \& Astrophysics}, , \vol{369}, 574.

Yoon, S.-C., Langer, N., Norman, C. 2006, \journal{Astronomy \& Astrophysics}, \vol{460,} 199.

\endreferences

\end{document}